\documentclass[10pt]{amsart}
\usepackage{amssymb}
\usepackage{enumerate}
\usepackage{epsf}
\usepackage{psfrag}
\DeclareGraphicsExtensions{.eps,.art,.ART,.ps}
\newcommand{\bbR}{\mathbb{R}}      

\newtheorem{Thm}{Theorem}[section]

\begin{document}
\begin{abstract}
We derive the covariant optimality conditions for rocket trajectories in general relativity, with and without a bound on the magnitude of the proper acceleration. The resulting theory is then applied to solve two specific problems: the minimum fuel consumption transfer between two galaxies in a FLRW model, and between two stable circular orbits in the Schwarzschild spacetime.
\end{abstract}
%
%
\title{The rocket problem in general relativity}
\author{Pedro G. Henriques and Jos\'{e} Nat\'{a}rio}
\address{Centro de An\'alise Matem\'atica, Geometria e Sistemas Din\^amicos, Departamento de Matem\'atica, Instituto Superior T\'ecnico, 1049-001 Lisboa, Portugal}
\thanks{Partially supported by FCT (Portugal).}
\maketitle
%
%
%
\section{Introduction}
The theory of optimal rocket trajectories in the context of Newtonian mechanics is beautifully developed in the classical reference \cite{Lawden63}. In this paper, we generalize this theory to the general relativity setting. We find that it is possible to formulate the theory in a fully covariant fashion, with all relevant auxiliary variables having geometric meaning.

While it can be argued that this generalization is of little practical interest (we are quite far from being able to build a rocket capable of moving at relativistic speeds!), it turns out that the geometric perspective of general relativity brings fresh insights into this classical application of control theory -- for instance, the relation of the primer equation with the Jacobi equation (Theorem~\ref{conditions}), or the fact that ignorable coordinates restrict variations to geodesics with the same Killing conserved quantities (Theorem~\ref{ignorable}). Moreover, this theory can be used to address theoretical issues in general relativity \cite{Natario12}.

The structure of the paper is as follows. In Section~\ref{section1}, we derive the well known rocket equation in general relativity, relating the initial and final rest masses of a rocket following a given spacetime trajectory. This is done mainly for the reader's convenience, as well as to fix the notation. In Section~\ref{section2}, we recall the details of Mayer problem in optimal control theory, which we use in Section~\ref{section3} to derive the optimality conditions under the assumption that the magnitude of the proper acceleration is bounded. These conditions are essentially a covariant differential equation for an auxiliary vector field defined along the trajectory, called the primer, whose direction gives the direction of the acceleration, and whose magnitude determines whether the rocket is undergoing maximum acceleration, intermediate acceleration, or free-falling. In Section~\ref{section4}, we examine the behavior of the optimal trajectories as we let the bound on the acceleration tend to infinity. We find that the maximum acceleration arcs collapse to instantaneous accelerations, whose locations and directions are again determined by the primer. In Section~\ref{section5}, we study the general $1$-dimensional problem, which we then particularize in Section~\ref{section6} to the problem of finding the minimum fuel consumption transfer between two galaxies in a FLRW model. Besides the general theory, we use a Lorentzian version of the Gauss-Bonnet Theorem to find these optimal trajectories, under the assumption that the expansion of the universe is either accelerating or non-accelerating. In Section~\ref{section7}, we study the role of ignorable coordinates, which we use in Section~\ref{section8} to solve the problem of finding the minimum fuel consumption transfer between two stable circular orbits in the Schwarzschild spacetime.

We use the conventions of \cite{MTW73}, including geometrized units (in which the speed of light and Newton's gravitational constant are set equal to $1$) and the Einstein summation convention; when using the latter, Greek letters will always represent spacetime indices.
%
%
\section{Rocket Equation in General Relativity}\label{section1}
To study a relativistic rocket moving in $d$ space dimensions, we must consider a $(d+1)$-dimensional spacetime $(M,g)$, that is, a smooth\footnote{By smooth we mean $C^\infty$.} $(d+1)$-dimensional manifold $M$ endowed with a Lorentzian metric $g$. We assume $(M,g)$ to be time-orientable, with a definite choice of time orientation. If the rocket's size is sufficiently small (as compared to the local radius of curvature), so that it can be modelled as a point particle, then its spacetime trajectory is given by a smooth future-directed timelike curve $c:[\tau_0,\tau_1] \subset \bbR \to M$, which we assume to be parameterized by the proper time $\tau$ (that is, the spacetime velocity $\dot{c}$ is a unit vector, $g(\dot{c},\dot{c})=-1$). Since the rocket must eject particles to accelerate, its rest mass will not be constant, but will be given instead by a smooth function $m:[\tau_0,\tau_1] \to \bbR^+$. For the same reason, the rocket's energy-momentum vector $m \dot{c}$ will not be conserved, and so we will have
\begin{equation}\label{J1}
\nabla_{\dot{c}} \left(m\dot{c}\right) + J = 0,
\end{equation}
where $\nabla$ is the Levi-Civita connection of $g$ and $J$ is a vector field along $c$ representing the instantaneous rate (with respect to proper time) at which energy-momentum is being carried away by the exhaust. The only {\em a priori} restriction to $J$ is that it must be timelike or null and future-directed, corresponding to the physical requirement that the speed of the exhaust particles must not exceed the speed of light. Rewriting \eqref{J1} as
\begin{equation}\label{J2}
J = - \dot{m} \dot{c} - m \nabla_{\dot{c}} \dot{c},
\end{equation}
and noting that the covariant acceleration $\nabla_{\dot{c}} \dot{c}$ is orthogonal to the unit tangent vector $\dot{c}$ (hence spacelike), we see that the {\em a priori} restriction on $J$ requires that $\dot{m} \leq 0$ (so that $J$ is future-directed), that is, the rocket's rest mass is a nonincreasing function. Moreover, we must have $J=0$ whenever $\dot{m}=0$ (because $J$ cannot be spacelike), and so
\begin{equation}\label{J3}
J = - \dot{m} (\dot{c} + V)
\end{equation}
for some spacelike vector field $V$ orthogonal to $\dot{c}$ satisfying $g(V,V) \leq 1$ (so that $J$ is timelike or null). Note that the norm $|V|:=|g(V,V)|^\frac12$ can be interpreted as the instantaneous speed of the exhaust particles as measured by the rocket. Comparing \eqref{J2} and \eqref{J3}, we have
\begin{equation}\label{dotm}
m \nabla_{\dot{c}} \dot{c} = \dot{m} V \Rightarrow \dot{m}|V| = - m |\nabla_{\dot{c}} \dot{c}|
\end{equation}
(the minus sign coming from the fact that $m$ is nonincreasing). 

In this paper, we will be interested in the case where speed of the exhaust particles is constant, that is, $|V|=v$ for some constant $v$ satisfying $0 < v \leq 1$ (the extreme case $v=1$ corresponding to the exhaust particles moving at the speed of light, that is, a photon rocket). Integrating \eqref{dotm} we then obtain the following result (see also \cite{Ackeret46, SMS47, Bade53, Rindler60, Burcev62, Rhee65, Pomeranz66}).

\begin{Thm}
For the relativistic rocket model specified above, the final rest mass $m_1:=m(\tau_1)$ is related to its initial rest mass $m_0:=m(\tau_0)$ by the formula
\begin{equation} \label{rocket}
m_1 = m_0 \exp\left(-\frac1v \int_{\tau_0}^{\tau_1} |\nabla_{\dot{c}}\,\dot{c}\,| d\tau\right).
\end{equation}
\end{Thm}

To minimize the fuel consumption, one must then minimize the action determined by the second order Lagrangian $L:=|\nabla_{\dot{c}}\,\dot{c}\,|$, which is not differentiable along its zero level set. For this reason, the standard Euler-Lagrange equations cannot be used directly, and one must turn to optimal control theory.
%
%
\section{Mayer Problem}\label{section2}
In this section, we briefly review the Mayer problem in optimal control. For more details see \cite{Lawden63, Mesterton09} and references therein.

The setup of the Mayer problem is as follows: on a compact interval $[\tau_0,\tau_1]$ we consider $n$ continuous, piecewise smooth\footnote{We take piecewise smooth functions on $[\tau_0,\tau_1]$ to be given by restrictions of smooth functions defined on $\bbR$ to the subintervals of a partition of $[\tau_0,\tau_1]$; in particular, piecewise smooth functions and all their derivatives have one-sided limits at all points.} functions $x^i:[\tau_0,\tau_1]\to\bbR$, called the {\bf state variables}, and $m$ piecewise smooth functions $u^j:[\tau_0,\tau_1]\to\bbR$, called the {\bf control variables}. Each choice of the control variables completely determines the state variables via the Cauchy problem
\[
\begin{cases}
x^i(\tau_0)=x^i_0, \\
\dot{x}^i(\tau)=f^i(x^1(\tau), \ldots, x^n(\tau), u^1(\tau), \ldots, u^m(\tau), \tau),
\end{cases}
\]
(where the functions $f^i$ are smooth\footnote{It is possible to formulate the Mayer problem with much lower regularity; here, for simplicity, we consider only (piecewise) smooth functions, as is usual in differential geometry.}). We require the control variables to satisfy the $p<m$ independent\footnote{That is, the Jacobian matrix $\left(\frac{\partial g^k}{\partial u^j}\right)$ has rank $p$.} constraint equations
\[
g^k(x^1(\tau), \ldots, x^n(\tau), u^1(\tau), \ldots, u^m(\tau), \tau) = 0
\]
for each $\tau \in [\tau_0,\tau_1]$, and the state variables to satisfy the endpoint conditions
\[
x^i(\tau_1)=x^i_1,
\]
for $i=1, \ldots, q \leq n$. The Mayer problem consists in choosing the control variables so that $J(x^{q+1}_1, \ldots, x^n_1)$ is minimized, where $J:\bbR^{n-q}\to\bbR$ is a given smooth function and $x^i_1:=x^i(\tau_1)$ for $i=q+1, \ldots, n$. The problem can be slightly generalized by letting $\tau_1$ be variable, in which case $J$ may also depend on $\tau_1$.

The solution of the Mayer problem is obtained from the {\bf Pontryagin Maximum Principle} as follows: we introduce $n$ additional variables $p_i$, called the {\bf momenta}, and define the {\bf Hamiltonian}
\[
H(x^1, \ldots, x^n, p_1, \ldots, p_n, \tau) := \max_{(u^1, \ldots, u^m)} p_i f^i(x^1, \ldots, x^n, u^1, \ldots, u^m, \tau),
\]
where the maximum is taken over all $u^j$ satisfying the constraint equations at time $\tau$. Then $J$ is minimized for state variables solving {\bf Hamilton's equations}
\[
\begin{cases}
\dot{x}^i=\frac{\partial H}{\partial p_i}, \\
\dot{p}_i=-\frac{\partial H}{\partial x^i},
\end{cases}
\]
with the initial conditions $p_i(\tau_0)$ chosen such that $x^i(\tau_1)=x^i_1$ for $i=1, \ldots, q$ and
\[
p_i(\tau_1)= - \frac{\partial J}{\partial x^i_1}
\]
for $i=q+1, \ldots, n$. If $\tau_1$ is allowed to vary, then its value can be determined from the additional endpoint condition
\[
H(\tau_1) = \frac{\partial J}{\partial \tau_1}.
\]

Computing the Hamiltonian is a constrained maximization problem, and leads to the conditions
\[
\frac{\partial F}{\partial u^j} = 0,
\]
where $F$ is the {\bf Lagrange expression}
\begin{align*}
& F(x^1, \ldots, x^n, p_1, \ldots, p_n, u^1, \ldots, u^m, \mu_1, \ldots, \mu_p,\tau) := \\
& \quad \quad p_i f^i(x^1, \ldots, x^n, u^1, \ldots, u^m, \tau) - \mu_k g^k(x^1, \ldots, x^n, u^1, \ldots, u^m, \tau).
\end{align*}
These conditions, together with the constraint equations, yield $u^j$ and $\mu_k$ as functions of the variables $(x^1, \ldots, x^n, p_1, \ldots, p_n, \tau)$, thereby introducing $p$ additional piecewise smooth functions $\mu_k:[\tau_0,\tau_1] \to \bbR$, which are called the {\bf Lagrange multipliers}.\footnote{The multiplier $\mu_k$ gives the rate of change of the Hamiltonian as the value of the constraint $g^k$ is varied.}

It is possible to use the Lagrange expression to obtain a shortcut for writing Hamilton's equations. Indeed, if the constraints are satisfied, then
\[
\frac{\partial g^k}{\partial x^l} + \frac{\partial g^k}{\partial u^j}\frac{\partial u^j}{\partial x^l} = 0,
\]
and hence
\begin{align*}
\frac{\partial H}{\partial x^l} & = p_i \frac{\partial f^i}{\partial x^l} + p_i \frac{\partial f^i}{\partial u^j}\frac{\partial u^j}{\partial x^l} = p_i \frac{\partial f^i}{\partial x^l} + \mu_k \frac{\partial g^k}{\partial u^j}\frac{\partial u^j}{\partial x^l} \\
& = p_i \frac{\partial f^i}{\partial x^l} - \mu_k \frac{\partial g^k}{\partial x^l} = \frac{\partial F}{\partial x^l}.
\end{align*}
Therefore, Hamilton's equations can be written as
\[
\begin{cases}
\dot{x}^i=\frac{\partial F}{\partial p_i}, \\
\dot{p}_i=-\frac{\partial F}{\partial x^i},
\end{cases}
\]
without the need to explicitly compute the Hamiltonian $H$.
%
%
\section{Optimality Conditions}\label{section3}
We now set up the problem of finding minimum fuel consumption trajectories as a Mayer problem. For that, we choose local coordinates $(x^0, x^1, \ldots, x^d)$ on our spacetime manifold $M$. In this coordinate system, the metric is given by some matrix $g_{\mu\nu}=g_{\mu\nu}(x^0, x^1, \ldots, x^d)$, and the Levi-Civita connection is determined by the Christoffel symbols $\Gamma^\mu_{\alpha\beta}=\Gamma^\mu_{\alpha\beta}(x^0, x^1, \ldots, x^d)$, which can be computed from the metric by the formula
\begin{equation} \label{Christoffel}
\Gamma^\mu_{\alpha \beta} = \frac12 g^{\mu\nu} \left( \partial_\alpha g_{\beta \nu} + \partial_\beta g_{\alpha \nu} - \partial_\nu g_{\alpha \beta}\right), \qquad (g^{\mu\nu}):=(g_{\mu\nu})^{-1}.
\end{equation}
If the rocket's trajectory $c:[\tau_0,\tau_1] \to M$ is given in these local coordinates by $x^\mu=x^\mu(\tau)$, then its spacetime velocity $U:=\dot{c}$ has components $U^\mu = \dot{x}^\mu$, and its covariant acceleration $A := \nabla_{\dot{c}}\dot{c} = \nabla_U U$ has components
\begin{equation} \label{covA}
A^\mu = \dot{U}^\mu + \Gamma^\mu_{\alpha\beta} U^\alpha U^\beta.
\end{equation}
As we have seen, the covariant acceleration $A$ is orthogonal to the spacetime velocity $U$, hence spacelike. Consequently, $A = a N$, where $a := |A|$ and $N$ is a unit spacelike vector (that is, $g(N,N)=1$). The reasonable assumption is that the rocket is able to control the direction and intensity of its exhaust, hence $N$ and $a$. We therefore choose $a$ and the components $N^\mu$ of $N$ as our control variables. Moreover, it is physically natural (and, as it turns out, mathematically desirable) to assume an upper bound $\bar{a}$ on the acceleration that the rocket is able to achieve. This leads to the constraint $a \leq \bar{a}$, which we will implement by introducing an extra control variable $\alpha$. Therefore, we have $d+3$ control variables $(a, N^0, N^1, \ldots, N^d, \alpha)$.

As state variables, describing the rocket's trajectory, we must then choose the rocket's spacetime coordinates $x^\mu$ and the components $U^\mu$ of its spacetime velocity. Moreover, we introduce an additional variable $I$ (called the {\bf cost}), which we will use to measure the fuel consumption. Therefore, we have $2d+3$ state variables $(x^0, x^1, \ldots, x^d, U^0, U^1, \ldots, U^d, I)$. From \eqref{covA}, the evolution of the first $2d+2$ state variables will be given by the equations
\[
\begin{cases}
\dot{x}^\mu = U^\mu, \\
\dot{U}^\mu + \Gamma^\mu_{\alpha\beta} U^\alpha U^\beta = a N^\mu,
\end{cases}
\]
and we choose
\[
\dot{I} = a.
\]
By \eqref{rocket}, the final mass of the rocket is will then be
\[
m_1 = m_0 \exp\left(-\frac{I_1}{v} \right),
\]
where $I_1:=I(\tau_1)$. Therefore, minimizing the fuel consumption is equivalent to minimizing the final cost, $J=I_1$. We will, however, keep $J$ general, as it makes little difference in the minimum conditions. 

The fact that we want to use the proper time $\tau$ as the time parameter must be specified through the constraint equation
\begin{equation}\label{constraint1}
g_{\mu\nu} U^\mu N^\nu = 0.
\end{equation}
This guarantees that $A=\nabla_U U$ is orthogonal to $U$, implying that the length of $U$ is constant; we will assume the initial conditions to be chosen so that
\begin{equation}\label{Uunit}
g_{\mu\nu} U^\mu U^\nu = -1
\end{equation}
(note that this is {\bf not} an extra constraint\footnote{We cannot use \eqref{Uunit} directly as the constraint because its Jacobian matrix, with respect to the control variables, vanishes.}). A further constraint is necessary to specify that $a$ is the norm of the acceleration, or, equivalently, that $N$ is a unit vector, namely
\begin{equation}\label{constraint2}
g_{\mu \nu} N^\mu N^\nu = 1.
\end{equation}
Finally, it is necessary to require that $0 \leq a \leq \bar{a}$. These inequalities can be accommodated by the constraint equation
\[
a(\bar{a}-a)=\alpha^2.
\]
We can now write the minimum conditions. The Lagrange expression is
\begin{align}\label{Lagrange}
F & = p_\mu\left(-\Gamma^\mu_{\alpha\beta} U^\alpha U^\beta + a N^\mu\right) + p_{\mu+d+1} U^\mu + p_{2d+2} \, a \\
\nonumber & - \mu_1 g_{\mu\nu} U^\mu N^\nu - \mu_2 \left(g_{\mu\nu} N^\mu N^\nu - 1\right) - \mu_3\left[a(\bar{a}-a)-\alpha^2\right].
\end{align}
Therefore, we have the Hamilton equations
\begin{equation}\label{Hamilton}
\begin{cases}
\dot{p}_\mu = - \frac{\partial F}{\partial U^\mu} = 2p_\gamma \Gamma^\gamma_{\alpha\mu} U^\alpha - p_{\mu+d+1} + \mu_1 g_{\mu\alpha} N^\alpha, \\
\dot{p}_{\mu+d+1} = - \frac{\partial F}{\partial x^\mu} = p_\gamma\partial_\mu\Gamma^\gamma_{\alpha\beta} U^\alpha U^\beta + \partial_\mu g_{\alpha\beta} \left( \mu_1 U^\alpha N^\beta + \mu_2 N^\alpha N^\beta \right), \\
\dot{p}_{2d+2} = - \frac{\partial F}{\partial I} = 0,
\end{cases}
\end{equation}
and the Pontryagin conditions
\begin{equation}\label{Pontryagin}
\begin{cases}
0 = \frac{\partial F}{\partial N^\mu} = p_\mu a - \mu_1 g_{\mu \nu} U^\nu - 2 \mu_2 g_{\mu \nu} N^\nu, \\
0 = \frac{\partial F}{\partial a} = p_\mu N^\mu + p_{2d+2} + \mu_3(2a-\bar{a}), \\
0 = \frac{\partial F}{\partial \alpha} = 2\mu_3 \alpha.
\end{cases}
\end{equation}
The last Pontryagin condition \eqref{Pontryagin} implies that either $\alpha=0$ (in which case we must have $a=0$ or $a=\bar{a}$) or $\mu_3=0$ (in which case we may have $0<a<\bar{a}$). Therefore, the rocket's trajectory is naturally divided into zero acceleration (ZA) arcs ($a=0$), maximal acceleration (MA) arcs ($a=\bar{a}$) and intermediate acceleration (IA) arcs ($0<a<\bar{a}$).\footnote{If we are minimizing the fuel consumption, then the IA arcs, where the acceleration is not constrained, must be solutions of the Euler-Lagrange equations for the second order Lagrangian
\[
L(x^0, \ldots, x^d,\dot{x}^0, \ldots, \dot{x}^d,\ddot{x}^0, \ldots, \ddot{x}^d) = \left[ g_{\mu\nu} \left(\ddot{x}^\mu + \Gamma^\mu_{\alpha\beta} \dot{x}^\alpha \dot{x}^\beta\right)\left(\ddot{x}^\nu + \Gamma^\nu_{\gamma\delta} \dot{x}^\gamma \dot{x}^\delta\right)  \right]^{\frac12},
\] 
subject to the constraint $g_{\mu\nu}\dot{x}^\mu\dot{x}^\nu=-1$.}

Assume first that $a > 0$. Then the first Pontryagin condition \eqref{Pontryagin}, together with \eqref{constraint1}, \eqref{Uunit} and \eqref{constraint2}, implies
\begin{equation}\label{momenta}
p_\mu + \left(p_\alpha U^\alpha\right) U_\mu = \rho N_\mu,
\end{equation}
where
\[
\rho := p_\mu N^\mu.
\]
Since $N$ has to be chosen so that Lagrange expression \eqref{Lagrange} is maximized, we must have $\rho \geq 0$ (because otherwise it would be possible to increase $\rho$, and so $F$, by reversing $N$). If the constraints are satisfied, the terms proportional to $a$ in the Lagrange expression are
\begin{equation}\label{k}
(\rho + p_{2d+2}) a =: ka.
\end{equation}
Since again the choice of $a$ has to maximize the Lagrange expression, we must have $k \geq 0$ on a MA arc (because otherwise it would be possible to increase $F$ by decreasing $a$) and $k = 0$ on a IA arc (because otherwise it would be possible to increase $F$ by either increasing or decreasing $a$).

On a ZA arc, the control variables $N^\mu$ can be chosen arbitrarily (as long as they satisfy the constraint equations). Therefore, we keep the choice
\[
N_\mu = \frac{P_\mu}{\rho},
\]
whenever $\rho\neq 0$, where
\[
P_\mu := p_\mu + \left(p_\alpha U^\alpha\right) U_\mu
\]
are the components of the projection of the vector $p$ on the hyperplane orthogonal to $U$ and
\[
\rho := \left( g^{\mu\nu} P_\mu P_\nu \right)^\frac12 = p_\mu N^\mu.
\]
With this choice, \eqref{momenta} and \eqref{k} still hold, and so maximizing the Lagrange expression with respect to to $a$ requires $k \leq 0$ on a ZA arc (because otherwise it would be possible to increase $F$ by increasing $a$). 

The Lagrange multipliers $\mu_1$ and $\mu_2$ can be determined by contracting the first Pontryagin condition \eqref{Pontryagin} with $U$ and $N$, which yields 
\begin{equation} \label{mu_1}
a p_\mu U^\mu + \mu_1 = 0 
\end{equation}
and
\begin{equation} \label{mu_2}
a\rho=2\mu_2.
\end{equation}

The first Hamilton equation \eqref{Hamilton} can be rewritten as
\begin{equation} \label{nablaUp}
\nabla_U p_\mu = - q_\mu + \mu_1 N_\mu,
\end{equation}
where
\begin{equation}\label{q}
q_\mu := p_{\mu+d+1} - \Gamma^\gamma_{\alpha\mu}p_\gamma U^\alpha.
\end{equation}
The second Hamilton equation \eqref{Hamilton} then yields
\begin{align}
\label{nablaUq} 
\hspace{-2cm} \nabla_U q_\mu  = & \dot{q}_\mu - \Gamma^\gamma_{\alpha\mu} q_\gamma U^\alpha \\
\nonumber = & \dot{p}_{\mu+d+1} - \partial_\beta \Gamma^\gamma_{\alpha\mu}p_\gamma U^\alpha U^\beta - \Gamma^\gamma_{\alpha\mu}\dot{p}_\gamma U^\alpha - \Gamma^\gamma_{\alpha\mu}p_\gamma \dot{U}^\alpha - \Gamma^\gamma_{\alpha\mu} q_\gamma U^\alpha \\
\nonumber = & \partial_\mu\Gamma^\gamma_{\alpha\beta} p_\gamma U^\alpha U^\beta + \partial_\mu g_{\alpha\beta} \left( \mu_1 g_{\alpha\beta} U^\alpha N^\beta + \mu_2 N^\alpha N^\beta \right) - \partial_\beta \Gamma^\gamma_{\alpha\mu}p_\gamma U^\alpha U^\beta \\
\nonumber & - \Gamma^\gamma_{\alpha\mu} \left( \Gamma^\delta_{\beta\gamma} p_\delta U^\beta - q_\gamma + \mu_1 N_\gamma \right) U^\alpha - \Gamma^\gamma_{\alpha\mu}p_\gamma \left( -\Gamma^\alpha_{\beta \delta} U^\beta U^\delta + a N^\alpha \right) - \Gamma^\gamma_{\alpha\mu} q_\gamma U^\alpha \\
\nonumber = & \left( \partial_\mu\Gamma^\gamma_{\alpha\beta} - \partial_\beta \Gamma^\gamma_{\alpha\mu} + \Gamma^\gamma_{\mu\delta} \Gamma^\delta_{\alpha\beta} - \Gamma^\gamma_{\beta\delta} \Gamma^\delta_{\alpha\mu} \right) p_\gamma U^\alpha U^\beta \\
\nonumber & + \partial_\mu g_{\alpha\beta} \left( \mu_1 U^\alpha N^\beta + \mu_2 N^\alpha N^\beta \right) - \mu_1 \Gamma^\gamma_{\alpha\mu} N_\gamma U^\alpha - \mu_1 \Gamma^\gamma_{\alpha\mu} U_\gamma N^\alpha - 2\mu_2 \Gamma^\gamma_{\alpha\mu} N_\gamma N^\alpha,
\end{align}
where we have used \eqref{momenta}, \eqref{mu_1} and \eqref{mu_2} to write
\[
a p_\gamma = - a \left( p_\delta U^\delta \right) U_\gamma + a \rho N_\gamma = \mu_1 U_\gamma + 2 \mu_2 N_\gamma.
\]
Now, using \eqref{Christoffel}, we have
\begin{align*}
& \partial_\mu g_{\alpha\beta} U^\alpha N^\beta  - \Gamma^\gamma_{\alpha\mu} N_\gamma U^\alpha - \Gamma^\gamma_{\alpha\mu} U_\gamma N^\alpha \\
& = \partial_\mu g_{\alpha\beta} U^\alpha N^\beta -  \frac12 g^{\gamma \beta} \left( \partial_\alpha g_{\beta \mu} + \partial_\mu g_{\beta \alpha} - \partial_\beta g_{\alpha \mu} \right) \left( N_\gamma U^\alpha + U_\gamma N^\alpha \right)  \\
& = \partial_\mu g_{\alpha\beta} U^\alpha N^\beta -  \frac12 \left( \partial_\mu g_{\alpha \beta} + \partial_\alpha g_{\beta \mu}  - \partial_\beta g_{\alpha \mu} \right) \left( U^\alpha N^\beta + N^\alpha U^\beta \right) \\
& = \partial_\mu g_{\alpha\beta} U^\alpha N^\beta -  \frac12 \partial_\mu g_{\alpha \beta} \left( U^\alpha N^\beta + N^\alpha U^\beta \right) = 0,
\end{align*}
and, similarly,
\begin{align*}
& \partial_\mu g_{\alpha\beta} N^\alpha N^\beta - 2 \Gamma^\gamma_{\alpha\mu} N_\gamma N^\alpha \\
& = \partial_\mu g_{\alpha\beta} N^\alpha N^\beta - g^{\gamma \beta} \left( \partial_\alpha g_{\beta\mu} + \partial_\mu g_{\beta\alpha} - \partial_\beta g_{\alpha\mu} \right) N_\gamma N^\alpha \\
& = \left( \partial_\mu g_{\alpha\beta} - \partial_\alpha g_{\beta\mu} - \partial_\mu g_{\beta\alpha} + \partial_\beta g_{\alpha\mu} \right) N^\alpha N^\beta = 0.
\end{align*}
Therefore, the terms proportional to $\mu_1$ and $\mu_2$ drop out of \eqref{nablaUq}, and we are left with
\[
\nabla_U q_\mu = R_{\mu \beta \,\,\, \alpha}^{\,\,\,\,\,\,\, \gamma} p_\gamma U^\alpha U^\beta = R_{\mu\alpha\beta\gamma}U^\alpha p^\beta U^\gamma = R_{\mu\alpha\beta\gamma}U^\alpha P^\beta U^\gamma.
\]

It is convenient to write the Hamilton equations using the modified momentum variables $P_\mu$. We have
\begin{align}
\label{nablaUP} \nabla_U P_\mu & = \nabla_U \left( p_\mu + \left(p_\alpha U^\alpha\right) U_\mu \right) = \nabla_U p_\mu + f U_\mu + \left(p_\alpha U^\alpha\right) a N_\mu \\
\nonumber & = - q_\mu + \mu_1 N_\mu + f U_\mu - \mu_1 N_\mu = - q_\mu + f U_\mu,
\end{align}
where we used \eqref{nablaUp} and \eqref{mu_1}, and defined
\[
f := \frac{d}{d\tau} \left(p_\alpha U^\alpha\right).
\]
This function can be determined from the Hamiltonian first integral
\[
H = p_\mu\left(-\Gamma^\mu_{\alpha\beta} U^\alpha U^\beta + a N^\mu\right) + p_{\mu+d+1} U^\mu + p_{2d+2} \, a,
\]
which, in view of \eqref{k} and \eqref{q}, can be written as
\[
H = q_\mu U^\mu + ka.
\]
In fact, if we contract \eqref{nablaUP} with $U$, we obtain
\begin{align*}
f = - \left(\nabla_U P_\mu\right) U^\mu - q_\mu U^\mu = P_\mu \left( \nabla_U U^\mu \right) - H + ka = \rho a + ka - H.
\end{align*}

If $\tau_1$ is not specified, then
\[
H = \frac{\partial J}{\partial \tau_1}.
\]
The remaining endpoint conditions depend on the exact nature of the problem. The standard problem of minimizing the fuel consumption, as we have seen, corresponds to $J = I_1$. In this case, the values of all state variables except $I$ are fixed at $\tau=\tau_1$, and we have the additional endpoint condition
\[
p_{2d+2} = - \frac{\partial J}{\partial I_1} = -1.
\]
Moreover, if $\tau_1$ is not fixed, then $H=0$.

Following \cite{Lawden63}, we shall call the spacelike vector field $P$ and the function $k$ the {\bf primer} and the {\bf switching function} of the trajectory. Like all variables in our problem, they are piecewise smooth. Moreover, it is known from the general theory of the Mayer problem that both the momentum variables and the Hamiltonian must be continuous along the optimal trajectory. Therefore, both the primer and the switching function are continuous. The continuity of $H$ then implies that $k$ must vanish at the discontinuities of $a$.

To summarize, we have proved the following result.

\begin{Thm}\label{conditions}
The optimal motion of the rocket is obtained from the motion equation 
\[
\nabla_U U^\mu = a N^\mu
\]
by choosing
\[
N^\mu = \frac{P^\mu}{\rho},
\]
where 
\[
\rho := \left( g^{\mu\nu} P_\mu P_\nu \right)^\frac12,
\]
and the primer $P^\mu$ satisfies
\begin{equation}\label{Hamilton2}
\begin{cases}
\nabla_U P_\mu = - q_\mu + (\rho a + ka - H) U_\mu, \\
\nabla_U q_\mu = R_{\mu\alpha\beta\gamma}U^\alpha P^\beta U^\gamma
\end{cases}
\end{equation}
and 
\[
P_\mu U^\mu  = 0.
\]
Here, the switching function $k := \rho + p_{2d+2}$, with $p_{2d+2}$ constant, satisfies $k \geq 0$ on MA arcs, $k=0$ on IA arcs and $k \leq 0$ on ZA arcs, and the Hamiltonian integration constant $H = q_\mu U^\mu + ka$ is given by
\[
H = \frac{\partial J}{\partial \tau_1}
\]
if $\tau_1$ is not fixed. Moreover, $p_{2d+2}=-1$ for the fuel minimizing problem.
\end{Thm}

%
%
\section{Instantaneous Accelerations}\label{section4}
The problem is simplified if we take the limit $\bar{a} \to + \infty$, corresponding to situations where the duration of the maximum acceleration periods is negligible. We assume that this limit exists and that all functions converge to piecewise smooth functions. In this limit, the MA arcs collapse to points where we have instantaneous (Dirac delta) accelerations.\footnote{See \cite{AKP11} for a general framework dealing with impulsive controls.} At such points we have:
\begin{enumerate}[(a)]
\item
$x^\mu$ is continuous, because $\frac{dx^\mu}{d\tau} = U^\mu$ is at most discontinuous;
\item
$q^\mu$ is continuous, because $\nabla_U q_\mu = R_{\mu\alpha\beta\gamma}U^\alpha P^\beta U^\gamma$ is at most discontinuous;
\item \label{last}
$U^\mu$ and $P^\mu$ have a discontinuity given by a (positive) boost in their plane, because $P_\mu U^\mu = 0$, $\nabla_U U^\mu$ is a positive multiple of $P^\mu$, and, in the limit, $\nabla_U P^\mu$ is a positive multiple of $U^\mu$ (in particular, $\rho$, and hence $k$, are continuous);
\item 
$k=0$, because this always happens at the endpoints of a MA arc.
\end{enumerate}

Both $k$ and $\rho$ attain their maximum at the instantaneous acceleration, where we must therefore have $\rho>0$. Statement (\ref{last}) can then be written as
\begin{equation} \label{boost}
\begin{cases}
U^\mu_+ =  U^\mu_- \cosh u + \frac{1}{\rho} P^\mu_- \sinh u, \\
P^\mu_+ = \rho U^\mu_- \sinh u + P^\mu_- \cosh u,
\end{cases}
\end{equation}
for some $u > 0$, where the indices $-$ and $+$ denote the limits at the instantaneous acceleration from the left and from the right. Since $ka=0$ on both ZA and IA arcs, the Hamiltonian integration constant yields
\[
q_\mu U^\mu_- = q_\mu U^\mu_+ = H,
\]
which, together with \eqref{boost}, imply
\[
q_\mu P^\mu_+ = \rho H \frac{\cosh u - 1}{\sinh u} = - q_\mu P^\mu_-.
\]
Since by \eqref{Hamilton2}
\[
\rho\dot{\rho} = P^\mu \left( \nabla_U P_\mu \right) = - P^\mu q_\mu,
\]
we have
\begin{equation} \label{dotrho}
\dot{\rho}_- = H \frac{\cosh u - 1}{\sinh u} = - \dot{\rho}_+,
\end{equation}
that is, $\dot{\rho}$ reverses its sign at the instantaneous acceleration. 

If $H=0$ (as is the case if $\tau_1$ is not specified and $J$ does not depend on $\tau_1$), then \eqref{dotrho} implies that $\rho$ is continuously differentiable with $\dot{\rho}=0$ at an instantaneous acceleration.\footnote{Note that this does not have to be true for instantaneous accelerations at the endpoints of the optimal trajectory.} In fact, in this case, $\rho$ is actually a $C^2$ function: using $ka=0$ on ZA and IA arcs, whence $q_\mu U^\mu=0$, and $P_\mu U^\mu =0$, we have
\begin{align*}
\dot{\rho}^2 + \rho\ddot{\rho} & = \left( \nabla_U P^\mu \right) \left( \nabla_U P_\mu \right) + P^\mu \left( \nabla_U \nabla_U P_\mu \right) \\
& = \left( -q^\mu + \rho a U^\mu \right) \left( -q_\mu + \rho a U_\mu \right) + P^\mu \left( -\nabla_U q^\mu + \rho a \nabla_U U^\mu \right) \\
& = q^\mu q_\mu - \rho^2 a^2 - P^\mu R_{\mu\alpha\beta\gamma}U^\alpha P^\beta U^\gamma + \rho^2 a^2 = q^\mu q_\mu + \rho^2 K,
\end{align*}
where $K := - R_{\alpha\beta\gamma\delta} U^\alpha N^\beta U^\gamma N^\delta$ is the sectional curvature of the plane spanned by $U$ and $P$, and so $\ddot{\rho}$ is continuous at an instantaneous acceleration.

We collect all these results in the following statement.

\begin{Thm}\label{instantaneous}
At an instantaneous acceleration we have:
\begin{enumerate}[(a)]
\item
$U$ and $P$ have a discontinuity given by a (positive) boost in their plane;
\item
$\rho$ reaches its maximum and is continuous;
\item
$\dot{\rho}$ reverses sign;
\item
if $H=0$, then $\rho$ is a $C^2$ function.
\end{enumerate}
\end{Thm}
%
%
\section{$1$-Dimensional Problems}\label{section5}
We now consider the simplest case $d=1$, corresponding to problems where the rocket moves along a line. The spacetime $(M,g)$ is thus $2$-dimensional.

Because $\rho$ is constant on IA arcs, we have from \eqref{Hamilton2}
\[
P^\mu q_\mu = - P^\mu \nabla_U P_\mu = - \rho \dot{\rho} = 0.
\]
Since $\{ U, N \}$ is an orthonormal frame and $q_\mu U^\mu = H$, we must have
\[
q_\mu = - H U_\mu,
\]
and hence, from the Hamilton equations \eqref{Hamilton2},
\[
-HaN_\mu = R_{\mu\alpha\beta\gamma}U^\alpha P^\beta U^\gamma \Leftrightarrow Ha=\rho K
\]
(where $K$ is the sectional curvature). It immediately follows that, if $H=0$ (as will be the case if $\tau_1$ is not specified and $J$ does not depend on $\tau_1$), then IA arcs can only exist in a flat spacetime. Even if $H \neq 0$, IA arcs can only exist if the curvature has the same sign as $H$.

On ZA arcs, on the other hand, we have $\nabla_U U = 0$, i.e.~$U$ is parallel transported. Since $\{ U, N \}$ is an orthonormal frame, it follows that $N$ is also parallel transported. Using $P = \rho N$, the Hamilton equations \eqref{Hamilton2} reduce to
\[
\ddot{\rho} N_\mu = - R_{\mu\alpha\beta\gamma}U^\alpha P^\beta U^\gamma \Leftrightarrow \ddot{\rho} = K\rho.
\]

In summary, we have the following result.

\begin{Thm}\label{1dim}
If $(M,g)$ is $2$-dimensional, then:
\begin{enumerate}[(a)]
\item
$Ha=\rho K$ on IA arcs;
\item
$\ddot{\rho} = K\rho$ on ZA arcs.
\end{enumerate}
\end{Thm}

%
%
\section{FLRW Models}\label{section6}
Recall that the Friedmann-Lema\^\i tre-Robertson-Walker (FLRW) models are given in spherical coordinates $(t,r,\theta,\varphi)$ by metrics of the form
\[
ds^2 = - dt^2 + R^2(t) \left[dr^2 + {\Sigma_k}^2(r) \left(d\theta^2+\sin^2\theta d\varphi^2\right)\right] \qquad (k=-1,0,1),
\]
where $\Sigma_0(r):=r$ (flat universe), $\Sigma_1(r):=\sin r$ (spherical universe) and $\Sigma_{-1}(r):=\sinh r$ (hyperbolic universe). These metrics describe spatially homogeneous and isotropic universes, and are believed to provide a good model for the large scale structure of our own universe. The matter content of these universes is usually taken to be a pressureless perfect fluid of uniform density $\rho$, whose elements are fixed at constant spatial coordinates, and are interpreted as (the averaged version of) galaxies (or clusters of galaxies). The Einstein field equations then determine the time evolution of the scale factor $R(t)$ and the matter density $\rho(t)$, according to the Friedmann equations
\[
\begin{cases}
\displaystyle \frac{\dot{R}^2}{R^2} = \frac{2E}{R^3} + \frac{\Lambda}{3} - \frac{k}{R^2}, \\ \\
\displaystyle \rho = \frac{3E}{4\pi R^3}
\end{cases}
\]
(where $E\geq 0$ is a constant determining the matter density and $\Lambda$ is the cosmological constant). The flat Minkowski spacetime is a particular case, corresponding to $k=E=\Lambda=0$.

We consider the problem of transfer between two galaxies in a FLRW universe. Assuming the first galaxy to be placed at $r=0$, it is easily seen that the motion must occur along the line of constant $(\theta,\varphi)$ joining the two galaxies, and so this is naturally a $1$-dimensional problem. Our spacetime will then be the totally geodesic $2$-dimensional submanifold of constant $(\theta,\varphi)$, with metric
\[
ds^2 = - dt^2 + R^2(t) dr^2.
\]
Choosing the natural orthonormal coframe
\[
\omega^0 := dt, \quad \quad \omega^1 := R(t) dr,
\]
one readily obtains from the first Cartan structure equations,
\[
d \omega^\alpha + \omega^\alpha_{\,\,\,\beta} \wedge \omega^\beta = 0,
\]
the nonvanishing connection forms
\[
\omega^0_{\,\,\,1} = \omega^1_{\,\,\,0} = \dot{R}(t) dr,
\]
and hence the nonvanishing curvature forms
\[
\Omega^0_{\,\,\,1} = \Omega^1_{\,\,\,0} = d\omega^0_{\,\,\,1} = \frac{\ddot{R}(t)}{R(t)} \omega^0 \wedge \omega^1.
\]
Consequently, the sectional curvature is
\[
K = - R_{0101} = \frac{\ddot{R}(t)}{R(t)}.
\]

Assume that the rocket aims at minimizing the fuel necessary to take it from being at rest in the first galaxy (placed at $r=0$) at time $t=t_0$ to being at rest in the second galaxy (placed at $r=r_1$) at time $t=t_1$.\footnote{We assume $(t_1,r_1)$ to be in the chronological future of $(t_0,0)$.} For this problem, $H=0$, and consequently there are no IA arcs if $K \neq 0$ for $t_0<t<t_1$. 

If $K(t)>0$ (that is, if the expansion is accelerating) for $t_0<t<t_1$, then, from Theorem~\ref{1dim}, we have $\ddot{\rho} = K\rho > 0$ for $\rho \neq 0$, and so all critical points of $\rho$ are necessarily minima. We conclude that there cannot exist instantaneous accelerations except at the endpoints of the optimal trajectory, which must then be a ZA arc with two instantaneous accelerations at the endpoints. 

The same conclusion holds if $K(t)\leq 0$ (that is, if the expansion is not accelerating) for $t_0<t<t_1$, but the argument is more complicated. We start by proving a Lorentzian version of the Gauss-Bonnet Theorem suited to our problem. To do so, we set
\[
E_0 := \frac{\partial}{\partial t}, \quad \quad E_1 := \frac1{R(t)} \frac{\partial}{\partial r},
\]
and consider a second orthonormal frame
\[
F_0 := \cosh u \, E_0 + \sinh u \, E_1, \quad \quad F_1 := \sinh u \, E_0 + \cosh u \, E_1
\]
(where $u$ is the locally defined {\bf hyperbolic angle} from $E_0$ to $F_0$). If $c: [\tau_0, \tau_1] \to M$ is a timelike curve parameterized by its proper time, then from
\[
\nabla_{\dot{c}} E_{\alpha} = \omega^\beta_{\,\,\,\alpha} (\dot{c}) E_{\beta},
\]
one readily obtains
\[
\nabla_{\dot{c}} F_0 = \left( \dot{u} + \omega^0_{\,\,\,1} (\dot{c}) \right) F_1.
\]
If $F_0$ is parallel transported along $c$, then we have
\[
\dot{u} = - \omega^0_{\,\,\,1} (\dot{c}),
\]
and hence
\begin{equation} \label{Delta u}
u(c(\tau_1)) - u(c(\tau_0)) = - \int_c \omega^0_{\,\,\,1}.
\end{equation}
If we now write
\[
\dot{c}(\tau) = \cosh w(\tau) \, F_0 + \sinh w(\tau) \, F_1
\]
(so that the total hyperbolic angle from $E_0$ to $\dot{c}$ is $u+w$), then we have
\[
\nabla_{\dot{c}}{\dot{c}} = \dot{w} (\sinh w \, F_0 + \cosh w \, F_1),
\]
and, consequently,
\[
\left|\nabla_{\dot{c}}\,\dot{c}\,\right| = \left|\dot{w}\right|.
\]
The function $\dot{w}(\tau)$ is called the {\bf geodesic curvature} of $c$. The cost of $c$ is
\begin{equation}\label{costgreaterthan}
\int_{\tau_0}^{\tau_1} \left|\nabla_{\dot{c}}\,\dot{c}\,\right| d \tau =  \int_{\tau_0}^{\tau_1} \left|\dot{w}\right| d \tau \geq \left|w(\tau_1)-w(\tau_0)\right|,
\end{equation}
with equality if and only if $\dot{w}$ does not change sign. 

\begin{figure}[h!]
\begin{center}
\psfrag{E0}{$E_0$}
\psfrag{c+}{$c^+$}
\psfrag{c-}{$c^-$}
\epsfxsize=.7\textwidth
\leavevmode
\epsfbox{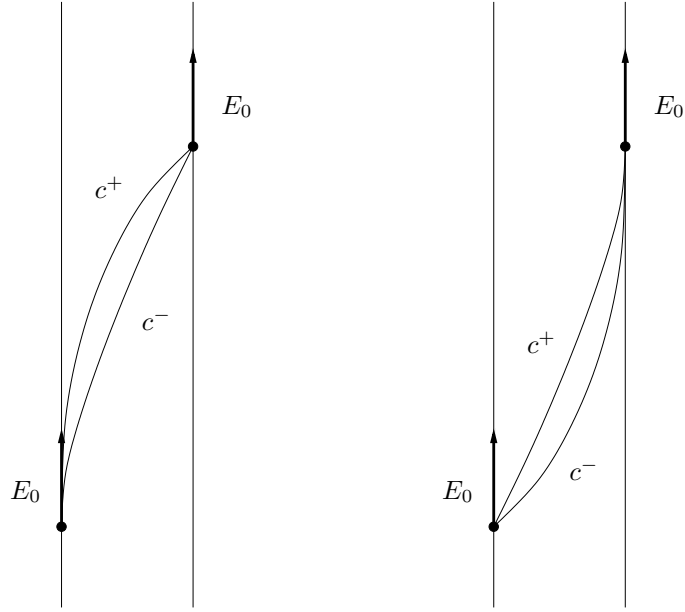}
\end{center}
\caption{Timelike curves tangent to $E_0$ at the initial or final endpoints.}\label{FLRW2}
\end{figure}

Now let $c^+$ and $c^-$ be timelike curves with common endpoints, with $c^+$ to the future of $c^-$. Assume the endpoints to be the only intersections and let $\Delta \subset M$ be the region bounded by the two curves. If the curves are tangent to $E_0$ at the initial endpoint (where we choose $F^+_0=F^-_0=E_0$), then we must have, at the final endpoint,
\[
w^+ + u^+ \geq w^- + u^-
\]
(cf.~Figure~\ref{FLRW2}). Now, from \eqref{Delta u}, we have
\[
u^+ - u^- = - \int_{c^+} \omega^0_{\,\,\,1} + \int_{c^-} \omega^0_{\,\,\,1} = \int_{\partial \Delta} \omega^0_{\,\,\,1} = \int_\Delta  d\omega^0_{\,\,\,1} = \int_\Delta  \Omega^0_{\,\,\,1} = \int_\Delta K \leq 0,
\]
implying that
\begin{equation} \label{+}
w^+ \geq w^-.
\end{equation}
Similarly, if the curves are tangent to $E_0$ at the final endpoint (where we choose $F^+_0=F^-_0=E_0$), then we must have, at the initial endpoint,
\[
w^- + u^- \geq w^+ + u^+
\]
(cf.~Figure~\ref{FLRW2}). Again, from \eqref{Delta u}, we have
\[
- u^+ + u^- = - \int_{c^+} \omega^0_{\,\,\,1} + \int_{c^-} \omega^0_{\,\,\,1} = \int_{\partial \Delta} \omega^0_{\,\,\,1} = \int_\Delta  d\omega^0_{\,\,\,1} = \int_\Delta  \Omega^0_{\,\,\,1} = \int_\Delta K \leq 0,
\]
implying that
\begin{equation} \label{-}
w^+ \leq w^-.
\end{equation}

\begin{figure}[h!]
\begin{center}
\psfrag{(t0,x0)}{$(t_0,0)$}
\psfrag{(t1,x1)}{$(t_1,r_1)$}
\psfrag{x=x0}{$r=0$}
\psfrag{x=x1}{$r=r_1$}
\psfrag{g}{$\gamma$}
\psfrag{c}{$c$}
\epsfxsize=.4\textwidth
\leavevmode
\epsfbox{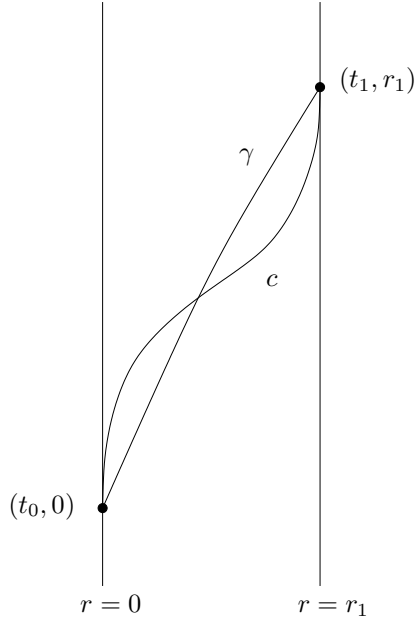}
\end{center}
\caption{Timelike curves connecting $(t_0,0)$ and $(t_1,r_1)$.}\label{FLRW}
\end{figure}

Finally, let $\gamma$ be the geodesic connecting the events $(t_0,0)$ and $(t_1,r_1)$, regarded as tangent to $E_0$ at the two endpoints via two instantaneous accelerations. Let $c$ be any smooth timelike curve with the same endpoints, also tangent to $E_0$ at the two endpoints. Then $c$ is to the future of $\gamma$ near $(t_0,0)$, $\gamma$ is to the future of $c$ near $(t_1,r_1)$ and the two curves intersect at least once besides the endpoints (cf.~Figure~\ref{FLRW}). If we take $c$ as $c^+$ and $\gamma$ as $c^-$ between the first endpoint and the first intersection (besides the endpoints), then we have from \eqref{+}
\[
\left|w^- - 0\right| = w^- \leq w^+ \leq \left|\omega^+ - 0\right|.
\]
Since the sign of $\dot{w}^-$ does not change, we see from \eqref{costgreaterthan} that the cost of $c$ is greater or equal than the cost of $\gamma$ between these two points. Analogously, if we take $\gamma$ as $c^+$ and $c$ as $c^-$ between final intersection (besides the endpoints) and the final endpoint, then we have from \eqref{-}
\[
\left|0 - w^+\right| = w^+ \leq w^- \leq \left|0 - w^-\right|.
\]
Since the sign of $\dot{w}^+$ does not change, we see  from \eqref{costgreaterthan} that the cost of $c$ is greater or equal than the cost of $\gamma$ between these two points. Since the cost of $\gamma$ between the first and the last intersections is zero, we conclude that the total cost of $c$ is greater or equal than the total cost of $\gamma$.

In fact, it is not difficult to see that the cost of $c$ is strictly greater than the cost of $\gamma$: if $c$ is tangent to $\gamma$ at any intersection (besides the endpoints), then its geodesic curvature must change signs. Therefore, we have the following result.

\begin{Thm}\label{FLRWThm}
If $\ddot{R}(t)>0$ or $\ddot{R}(t)\leq 0$ for $t_0 < t < t_1$, then the minimum fuel consumption trajectory for the transfer of a rocket from being at rest with respect to the galaxy $r=0$ at time $t=t_0$ to being at rest with respect to the galaxy $r=r_1$ at time $t=t_1$ is a ZA arc with instantaneous accelerations at the endpoints.
\end{Thm}

Note that the conditions $\ddot{R}(t)>0$ and $\ddot{R}(t)\leq 0$ are really the conditions $K>0$ and $K\leq 0$, and so this theorem is easily applicable to other $1$-dimensional problems.
%
%
\section{Ignorable Coordinates}\label{section7}
We say that a given coordinate $x^\sigma$ is {\bf ignorable} iff the metric does not depend on $x^\sigma$ and the final value $x^\sigma_1$ is not specified. Under certain circumstances, the existence of ignorable coordinates greatly simplifies the minimum conditions for ZA arcs.

Assume that $H=0$ (as is the case if $\tau_1$ is not specified and $J$ does not depend on $\tau_1$). In this case, the minimum equations on a ZA arc reduce to the geodesic and Jacobi (geodesic deviation) equations,
\[
\begin{cases}
\nabla_U U^\mu = 0, \\
\nabla_U \nabla_U P_\mu = - R_{\mu\alpha\beta\gamma}U^\alpha P^\beta U^\gamma,
\end{cases}
\]
and the Hamiltonian first integral becomes
\[
\left( \nabla_U P_\mu \right) U^\mu = 0,
\]
which is just the proper time derivative of the constraint $P_\mu U^\mu = 0$. Since the metric does not depend on the ignorable coordinate $x^\sigma$, the covariant component $U_\sigma$ is conserved. Moreover, the Hamiltonian is then independent of $x^\sigma$, and hence $p_{\sigma+d+1}$ is also constant. Since $x^\sigma_1$ is not fixed, we have $p_{\sigma+d+1} = 0$, that is,
\[
0 = p_{\sigma+d+1} = q_\sigma + \Gamma^\gamma_{\alpha\sigma}p_\gamma U^\alpha = - \nabla_U P_\sigma + \Gamma^\gamma_{\alpha\sigma}p_\gamma U^\alpha.
\]
On the other hand, we have
\[
0 = \nabla_U U_\sigma = \dot{U}_\sigma - \Gamma^\gamma_{\alpha\sigma} U_\gamma U^\alpha = - \Gamma^\gamma_{\alpha\sigma} U_\gamma U^\alpha,
\]
implying that
\[
\Gamma^\gamma_{\alpha\sigma}P_\gamma U^\alpha = \Gamma^\gamma_{\alpha\sigma}p_\gamma U^\alpha,
\]
and, consequently,
\begin{equation} \label{same_momenta}
\nabla_U P_\sigma - \Gamma^\gamma_{\alpha\sigma}P_\gamma U^\alpha = 0.
\end{equation}
We can obtain a geometrical interpretation of this condition by rewriting \eqref{same_momenta} as
\begin{equation} \label{same_momenta2}
X^\mu \nabla_U P_\mu - \left( \nabla_\alpha X^\gamma \right) P_\gamma U^\alpha= 0,
\end{equation}
where
\[
X = \frac{\partial}{\partial x^\sigma}.
\] 
Thinking of $P$ as a Jacobi field, hence satisfying
\[
\left[U,P\right] = 0 \Leftrightarrow \nabla_U P^\mu - \nabla_{P} U^\mu = 0,
\]
and using the fact the $X$ is a Killing field,
\[
\nabla_\mu X_\nu + \nabla_\nu X_\mu = 0,
\]
we can write \eqref{same_momenta} as
\[
X^\mu \nabla_{P} U_\mu + U^\mu \nabla_{P} X_\mu = 0 \Leftrightarrow \nabla_{P} \left( U_\mu X^\mu \right) = 0.
\]
Therefore, \eqref{same_momenta} is the condition that the Jacobi field $P^\mu$ connects infinitesimally nearby geodesics with the same value of the conserved quantity $U_\mu X^\mu$. In other words, we have the following result.

\begin{Thm}\label{ignorable}
If $H=0$ and $x^\sigma$ is an ignorable coordinate, then the Jacobi field $P$ connects infinitesimally nearby geodesics with the same value of the conserved quantity $U_\sigma$.
\end{Thm}
%
%
\section{Schwarzschild Solution}\label{section8}
Recall that the Schwarzschild solution is the unique spherically symmetric solution of the vacuum Einstein field equations, given in spherical coordinates $(t,r,\theta,\varphi)$ by the metric
\[
ds^2 = - \left( 1 - \frac{2M}r \right) dt^2 +  \left( 1 - \frac{2M}r \right)^{-1} dr^2 + r^2 \left(d\theta^2 + \sin^2 \theta d \varphi^2\right).
\]
It represents the gravitational field created by a spherically symmetric body of mass $M$ (which we assume to be positive), and so models the vicinity of an isolated planet or star (or even a more exotic object, like a neutron star or a black hole).

We consider the problem of finding the most fuel-efficient transfer between two stable circular orbits in the Schwarzschild metric. For simplicity, we restrict ourselves to motions the $(2+1)$-dimensional totally geodesic submanifold $\theta = \frac{\pi}2$, corresponding to the equatorial plane, whose metric is
\[
ds^2 = - \left( 1 - \frac{2M}r \right) dt^2 +  \left( 1 - \frac{2M}r \right)^{-1} dr^2 + r^2 d \varphi^2.
\]
We will not fix $\tau_1$ (hence $H=0$), $t_1$ or $\varphi_1$ (and so $t$ and $\varphi$ are ignorable coordinates). Moreover, the choice $J=I_1$ implies $k=\rho-1$. The geodesic equations in the coordinate system above can be written as
\begin{equation}\label{geodesic}
\begin{cases}
\displaystyle \ddot{t} + \frac{2M}{r^2}\left( 1 - \frac{2M}r \right)^{-1} \dot{t} \dot{r} = 0 \Leftrightarrow \left( 1 - \frac{2M}r \right) \dot{t} = E,\\ \\
\displaystyle \ddot{r} + \frac{M}{r^2}\left( 1 - \frac{2M}r \right) \dot{t}^2 - \frac{M}{r^2}\left( 1 - \frac{2M}r \right)^{-1} \dot{r}^2 - r \left( 1 - \frac{2M}r \right) \dot{\varphi}^2 = 0, \\ \\
\displaystyle \ddot{\varphi} + \frac2r \dot{r} \dot{\varphi} = 0 \Leftrightarrow r^2 \dot{\varphi} = L,
\end{cases}
\end{equation}
where $E$ and $L$ are integration constants. Equation \eqref{same_momenta} for $x^\sigma = t$ can be obtained by linearizing the first geodesic equation \eqref{geodesic} while keeping $E$ fixed:
\begin{equation} \label{dot_p_t}
\left( 1 - \frac{2M}r \right) \dot{P}^t + \frac{2M}{r^2} P^r \dot{t}  = 0 \Leftrightarrow \dot{P}^t + \frac{2M}{r^2} \dot{t} P_r = 0.
\end{equation}
Similarly, equation \eqref{same_momenta} for $x^\sigma = \varphi$ can be obtained by linearizing the third geodesic equation \eqref{geodesic} while keeping $L$ fixed:
\begin{equation} \label{dot_p_varphi}
r^2 \dot{P}^\varphi + 2r P^r \dot{\varphi} = 0 \Leftrightarrow \dot{P}^\varphi + \frac2r \left( 1 - \frac{2M}r \right) \dot{\varphi} P_r = 0.
\end{equation}
Since $U_t$ and $U_\varphi$ are conserved, the proper time derivative of the constraint
\[
P^t U_t + P^\varphi U_\varphi + P_r U^r = 0  
\]
is
\[
\dot{P}^t U_t + \dot{P}^\varphi U_\varphi + \frac{d}{d \tau} \left( P_r \dot{r} \right) = 0.  
\]
Using \eqref{dot_p_t} and \eqref{dot_p_varphi} yields
\begin{align*}
\frac{2M}{r^2}\left( 1 - \frac{2M}r \right) \dot{t}^2 P_r - 2r \left( 1 - \frac{2M}r \right) \dot{\varphi}^2 P_r + \frac{d}{d \tau} \left( P_r \dot{r} \right) = 0,
\end{align*}
and from the radial component of the geodesic equations \eqref{geodesic} we have
\begin{equation} \label{Prime_eq}
- 2\ddot{r} P_r + \frac{2M}{r^2}\left( 1 - \frac{2M}r \right)^{-1} \dot{r}^2 P_r + \dot{P}_r \dot{r} + P_r \ddot{r} = 0.
\end{equation}
For non-circular orbits, this equation is readily solved:
\[
P_r = A \dot{r} \left( 1 - \frac{2M}r \right)^{-1} \Leftrightarrow P^r = A \dot{r},
\]
where $A \in \bbR$ is an integration constant. Substituting on \eqref{dot_p_t} and using the first geodesic equation \eqref{geodesic} yields
\[
\dot{P}^t = A \ddot{t} \Leftrightarrow P^t = A \dot{t} + B,
\]
where $B \in \bbR$ is another integration constant. Similarly, substituting on \eqref{dot_p_varphi} and using the third geodesic equation \eqref{geodesic} yields
\[
\dot{P}^\varphi = A \ddot{\varphi} \Leftrightarrow P^\varphi = A \dot{\varphi} + C,
\]
where $C \in \bbR$ is yet another integration constant. Therefore, $P^\mu$ is a linear combination of three obvious Jacobi fields:
\[
P = A U + B \frac{\partial}{\partial t} + C \frac{\partial}{\partial \varphi}.
\]
The condition $P^\mu U_\mu = 0$ imposes a restriction on the integration constants, namely
\[
- A + B U_t + C U_\varphi = 0.
\]
We then see that
\begin{align*}
\rho^2 & = P_\mu P^\mu = - A^2 - \left( 1 - \frac{2M}r \right) B^2 + r^2 C^2 + 2ABU_t + 2ACU_\varphi \\
& = A^2 + \left( \frac{2M}r - 1 \right) B^2 + r^2 C^2,
\end{align*}
and, consequently,
\begin{equation} \label{noncircular}
2 \rho \dot{\rho} = \left( - \frac{2M}{r^2} B^2 + 2r C^2 \right) \dot{r}.
\end{equation}
Thus $\dot{\rho}$ can only be zero if $\dot{r}=0$ or $C^2 r^3 = M B^2$. This second possibility, however, corresponds to a minimum of $\rho^2$, and hence cannot happen at an instantaneous acceleration or transition to an IA arc. We conclude that the optimal trajectory can only enter or leave a non-circular orbit at a turning point, that is, a point where $\dot{r}=0$. At these points, we will have $P^r = A \dot{r} = 0$, meaning that the acceleration will be tangential to the orbit.

For circular orbits, \eqref{Prime_eq} does not yield any information, and we have to write the radial component of the Jacobi equation. A shortcut to do so is the following: using the normalization condition
\[
- \left( 1 - \frac{2M}r \right) \dot{t}^2 +  \left( 1 - \frac{2M}r \right)^{-1} \dot{r}^2 + r^2 \dot{\varphi}^2 = -1,
\]
we can recast the radial component of the geodesic equations \eqref{geodesic} in the form
\begin{equation}\label{geodesic_r}
\ddot{r} + \frac{M}{r^2} - \left( 1 - \frac{3M}r \right) \frac{L^2}{r^3} = 0.
\end{equation}
Linearizing the equation above while keeping $L$ constant yields
\begin{equation} \label{ddot{P^r}}
\ddot{P}^r + \left( 1 - \frac{6M}r \right) \frac{L^2}{r^4} P^r = 0,
\end{equation}
where we used \eqref{geodesic_r} to write
\[
M = \left( 1 - \frac{3M}r \right) \frac{L^2}{r}
\]
on a circular orbit. From
\[
P^\mu U_\mu = 0 \Leftrightarrow P^t U_t + P^\varphi U_\varphi = 0 \Leftrightarrow P^t = - \frac{U_\varphi}{U_t} P^\varphi 
\]
and
\[
\frac{U_\varphi}{U_t} = - \frac{r^2 \dot{\varphi}}{\left( 1 - \frac{2M}r \right) \dot{t}} = - \frac{\sqrt{Mr}}{1 - \frac{2M}r} 
\]
we have
\[
P^t = \frac{\sqrt{Mr}}{1 - \frac{2M}r} P^\varphi.
\]
Therefore
\begin{align*}
\rho^2 & = - \left( 1 - \frac{2M}r \right) \frac{Mr}{\left(1 - \frac{2M}r\right)^2} \left(P^\varphi\right)^2 + \left( 1 - \frac{2M}r \right)^{-1} \left(P^r\right)^2 + r^2 \left(P^\varphi\right)^2\\
& = \left( 1 - \frac{2M}r \right)^{-1} \left( \left(P^r\right)^2 + \left( 1 - \frac{3M}r \right) r^2 \left(P^\varphi\right)^2 \right),
\end{align*}
and hence
\begin{align*}
\rho \dot{\rho} & = \left( 1 - \frac{2M}r \right)^{-1} \left( P^r \dot{P}^r + \left( 1 - \frac{3M}r \right) r^2 P^\varphi \dot{P}^\varphi \right) \\
& = \left( 1 - \frac{2M}r \right)^{-1} P^r \left( \dot{P}^r - \left( 1 - \frac{3M}r \right) P^\varphi \frac{2L}{r} \right),
\end{align*}
where we used \eqref{dot_p_varphi}. So, for $\rho=1$, we will have $\dot{\rho}=0$ if $P^r = 0$, or
\begin{equation} \label{condition2}
\dot{P}^r = \left( 1 - \frac{3M}r \right) r^2 P^\varphi \frac{2L}{r^3}.
\end{equation}
In this second case we will have
\begin{align}
\ddot{\rho} & = \left( 1 - \frac{2M}r \right)^{-1} P^r \left( \ddot{P}^r - \left( 1 - \frac{3M}r \right) \dot{P}^\varphi \frac{2L}{r} \right) \nonumber \\
& = \left( 1 - \frac{2M}r \right)^{-1} P^r \left( - \left( 1 - \frac{6M}r \right) \frac{L^2}{r^4} P^r + \left( 1 - \frac{3M}r \right) \frac{4L^2}{r^4} P^r \right) \nonumber \\
& = \frac{3L^2}{r^4} \left( P^r \right)^2, \label{ddot_P}
\end{align}
which is strictly positive unless $P^r = 0$. We conclude that the optimal trajectory can only enter or leave a circular orbit at points where $P^r = 0$, meaning that the acceleration will be tangential to the orbit.

We now show that the optimal trajectory cannot contain IA arcs. Let us start by showing that, if a IA arc exists, then none of its endpoints can be a non-circular ZA arc. Since $\rho$ is a $C^2$ function and $\rho\equiv 1$ on a IA arc, then we must have $\ddot{\rho}=0$ at ZA arc's endpoint. From \eqref{noncircular} and the fact that $\dot{\rho}=\dot{r}=0$ at the endpoint, we obtain
\[
2 \rho \ddot{\rho} = \left( - \frac{2M}{r^2} B^2 + 2r C^2 \right) \ddot{r}.
\]
Now, $\ddot{r}$ never vanishes at a turning point, and the expression in brackets vanishes only at the minimum of $\rho$. We conclude that $\ddot{\rho}$ cannot vanish at the endpoint, and hence a IA arc cannot have an endpoint on a noncircular ZA arc.

For circular orbits, the story is more complicated. Using the formula
\[
\left( 1 - \frac{2M}r \right)^{-1} \left( \ddot{P}^r - \left( 1 - \frac{3M}r \right) \dot{P}^\varphi \frac{2L}{r} \right) = \frac{3L^2}{r^4} P^r,
\]
obtained while deducing \eqref{ddot_P}, we have
\[
\dot{\rho}^2 + \rho \ddot{\rho} = \left( 1 - \frac{2M}r \right)^{-1} \dot{P}^r \left( \dot{P}^r - \left( 1 - \frac{3M}r \right) P^\varphi \frac{2L}{r} \right) + \frac{3L^2}{r^4} \left( P^r \right)^2.
\]
Since on a transition point to a $IA$ arc we must have $\rho=1$ and $\dot{\rho}=\ddot{\rho}=P^r=0$, the formula above requires that either $\dot{P}^r=0$ or \eqref{condition2} holds. Differentiating again yields
\begin{align*}
& 3\dot{\rho}\ddot{\rho} + \rho\dddot{\rho} = \left( 1 - \frac{2M}r \right)^{-1} \ddot{P}^r \left( \dot{P}^r - \left( 1 - \frac{3M}r \right) P^\varphi \frac{2L}{r} \right) + \frac{9L^2}{r^4} P^r \dot{P}^r \\
& = - \left( 1 - \frac{2M}r \right)^{-1} \left( 1 - \frac{6M}r \right) \frac{L^2}{r^4} P^r  \left( \dot{P}^r - \left( 1 - \frac{3M}r \right) P^\varphi \frac{2L}{r} \right) + \frac{9L^2}{r^4} P^r \dot{P}^r,
\end{align*}
which shows that $\dddot{\rho}=0$ at the transition point. Finally, differentiating once more and assuming that \eqref{condition2} holds, we have
\[
\ddddot{\rho} = \frac{9L^2}{r^4} \left(\dot{P}^r\right)^2,
\]
showing that, if $\dot{P}^r\neq 0$, then $\rho$ would have a local minimum at the transition point. But that is impossible, since $\rho\leq 1$ on the circular orbit; therefore, we must have $P^r = \dot{P}^r=0$ at the transition point. Since $P^r$ satisfies the second-order linear ODE \eqref{ddot{P^r}}, we conclude that $P^r$ must vanish identically on the circular orbit, and hence $P^\mu$ must be a linear combination of the Killing vector fields $\frac{\partial}{\partial t}$ and $\frac{\partial}{\partial \varphi}$. These fields are orthogonal and have constant norm over the circular orbit, and hence we will have $\rho \equiv 1$ on the circular orbit.

We then see that a IA arc can only join circular orbits where $\rho \equiv 1$. The same reasoning as for IA arcs shows that these circular orbits cannot have an endpoint on a non-circular orbit. We conclude that, if the optimal trajectory contains a IA arc, then it must consist solely of IA arcs, circular orbits and instantaneous accelerations, and, moreover, we must have $\rho \equiv 1$ throughout the trajectory. 

On the other hand, if the optimal trajectory contains a stable circular orbit, then it is always possible to add a non-circular ZA arc. Indeed, if the trajectory departs from the stable circular orbit via a IA arc, then we can interrupt this arc after a sufficiently small proper time interval to achieve a non-circular orbit. This orbit is periodic in $r$ and will return to the  same values of $(r,\dot{r})$ after a certain proper time period, at which point the initial IA arc can be resumed (from different values of $t$ and $\varphi$, but the final values of these variables are not fixed). If there is an instantaneous acceleration at the stable circular orbit's endpoint, then it can be broken into an initial small enough instantaneous acceleration, followed by a non-circular ZA arc, followed by the remaining instantaneous acceleration (again from a point with different values of $t$ and $\varphi$). This has exactly the same cost as the original instantaneous acceleration, and so the resulting trajectory will also be optimal. We conclude that, if the optimal trajectory contains a stable circular orbit, then no IA arc can exist. 

Therefore we have proved the following result.

\begin{Thm}\label{Schwarzschild}
The optimal trajectory for the transfer of a rocket between two stable circular orbits in the Schwarzschild metric with minimum fuel consumption is composed of ZA arcs and instantaneous accelerations tangent to the ZA arcs, which must be at turning points for noncircular ZA arcs.
\end{Thm}

\begin{figure}[h!]
\begin{center}
\psfrag{ZA arc}{ZA arc}
\psfrag{Inst}{Instantaneous acceleration}
\epsfxsize=.6\textwidth
\leavevmode
\epsfbox{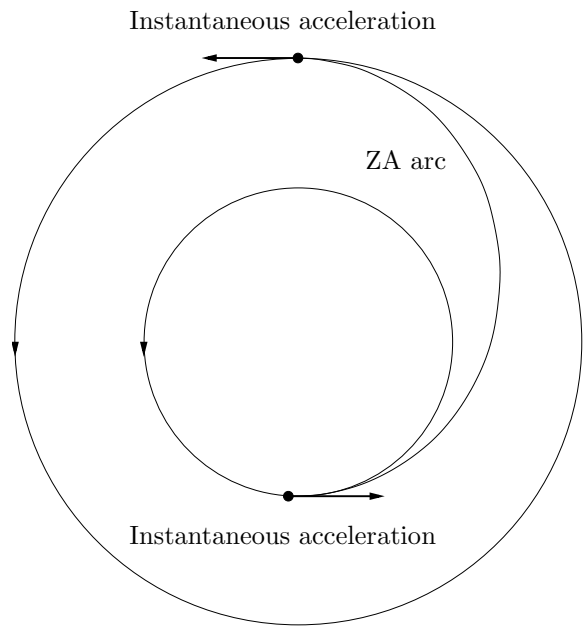}
\end{center}
\caption{Hohmann transfer manoeuvre.}\label{Hohmann}
\end{figure}

The simplest of such trajectories is the relativistic analogue of the celebrated {\bf Hohmann transfer manoeuvre} \cite{Hohmann25, Lawden63}, consisting of a single noncircular ZA arc connecting the two stable circular orbits (cf. Figure~\ref{Hohmann}).

%
%
\section{Conclusions}
We have derived the covariant optimality conditions for rocket trajectories in general relativity, with and without a bound on the magnitude of the proper acceleration (Theorems~\ref{conditions} and \ref{instantaneous}). These conditions are essentially a covariant differential equation for an auxiliary vector field defined along the trajectory (the primer), closely related to the Jacobi equation. We studied these conditions for general $1$-dimensional problems (Theorem~\ref{1dim}), and then particularized to the problem of the optimal transfer between two galaxies in a FLRW model. We found that for generic classes of models (including the flat Minkowski spacetime) the optimal trajectory consists of an initial instantaneous acceleration, followed by free-fall motion and a final instantaneous acceleration (Theorem~\ref{FLRWThm}). We showed that ignorable coordinates lead to great simplifications in the optimality conditions (Theorem~\ref{ignorable}), and used this to study the problem of optimal transfer between two stable circular orbits in the Schwarzschild spacetime. Here we found that the optimal trajectories consist of free fall motions connected by tangential instantaneous accelerations (Theorem~\ref{Schwarzschild}), including the relativistic analogue of the celebrated Hohmann transfer manoeuvre.
%
%

\end{document}